\documentclass{elsart}

\usepackage{natbib}
\usepackage{graphicx}
\usepackage{epsfig}

\usepackage{amssymb}

\begin{document}

\begin{frontmatter}



\title{Galaxy Colours in the  AKARI Deep SEP Survey}


\author[isas,esa]{Chris P. Pearson\corauthref{cor}}
\ead{cpp@ir.isas.jaxa.jp}
\author[isas]{Woong-Seob Jeong}
\author[isas]{Shuji Matsuura}
\author[isas]{Hideo Matsuhara}
\author[isas]{Takao Nakagawa}
\author[nagoya]{Hiroshi Shibai}
\author[nagoya]{Mitsunobu Kawada}
\author[isas]{Toshinobu Takagi}
\author[snu]{Hyung Mok Lee}
\author[isas]{Mai Shirahata}

\corauth[cor]{Corresponding author. Tel.: +81-42-759-8519; Fax:
+81-42-786-7202. \\ http://www.ir.isas.jaxa.jp/$\sim$cpp/astrof/}
\address[isas]{Institute of Space and Astronautical Science, Japan Aerospace
Exploration Agency, Yoshinodai 3-1-1, Sagamihara, Kanagawa 229-8510, Japan}
\address[esa]{ISO Data Centre, European Space Agency, Villafranca del Castillo, P.O. Box 50727, 28080 Madrid, Spain}
\address[snu]{Department of Physics and Astronomy, Seoul National University,
Shillim-Dong, Kwanak-Gu, Seoul 151-742, South Korea}
\address[nagoya]{Graduate School of Science, Nagoya University, Furo-cho,
Chigusa-gu, Nagoya 464-8602, Japan}

\begin{abstract}

 We investigate the segregation of the extragalactic population via colour criteria to produce an efficient and inexpensive methodology to select specific source populations as a function of far-infrared flux. Combining galaxy evolution scenarios and a detailed spectral library of galaxies, we  produce simulated catalogues incorporating segregation of the extragalactic population into component types (Normal, star-forming, AGN) via color cuts. As a practical application  we apply our criteria to the deepest survey to be undertaken in the far-infrared with the AKARI (formerly ASTRO-F) satellite. Using the far-infrared wavebands of the  Far-Infrared Surveyor (FIS, one of the focal-plane instruments on AKARI) we successfully segregate the normal, starburst and ULIRG populations. We also show that with additional MIR imaging from AKARI's Infrared Camera (IRC), significant contamination and/or degeneracy can be further decreased and show a particular example of the separation of cool normal galaxies and cold ULIRG sources. We conclude that our criteria provide an efficient means of selecting source populations (including rare luminous objects) and  produce colour-segregated source counts without the requirement of  time intensive ground-based follow up to differentiate between the general galaxy population.

\end{abstract}

\end{frontmatter}

\section{Introduction}
Fueled by rapid advances in detector technology, enabling fast and accurate
mapping of large areas of the sky, extragalactic astrophysics is enjoying a
golden age where astronomers have become spoiled for choice with the enormous
statistical survey data sets available probing out to cosmologically
significant distances such as the optical Sloan Digital Sky Survey (SDSS)
covering one quarter of the sky \citep{adelman-mccarthy06}, the near infrared
UKIRT Infrared Deep Sky Survey (UKIDSS, 4000 square degrees \citep{lawrence06})
and the \textit{Spitzer} Wide-area Infrared Extragalactic Survey (SWIRE,
\citet{lonsdale04}). The most basic tool for the study of cosmological galaxy
evolution is the monochromatic (single band) number count of extragalactic
sources. However aside from source counts and clustering analysis of large
scale structure, deeper investigation into the extragalactic population and
cosmological evolution will inevitable require multi-wavelength analysis and
hence galaxy colours. Ultimately we would like to define the extragalactic zoo
species by species and derive their corresponding distances to determine the
star-formation rate and sequence of structure formation in the Universe.
However such studies often involve extremely time intensive follow up campaigns with
either deep imaging or spectroscopy. As an intermediate step, galaxy colours
have been frequently used to gain a cheap yet effective insight into galaxy
populations and rough distances at a fraction of the cost of large scale follow-up campaigns. Such colour analysis has been
extremely effective in selecting and segregating populations of sources in the
near-infrared, optical and ultra-violet wavebands (e.g.  the Lyman break
population  \citep{steidel96, madau96},  ERO \&  \textit{BzK} sources
\citep{elston88,  cimatti02}). At infrared wavelengths astronomers have often
had to work with fewer bands and less well behaved detectors, yet the colour
criteria has also been applied with great success in the mid-far-infrared with the
pioneering \textit{IRAS} mission's all-sky survey  \citep{soifer87}. Using
simple colour criteria,  \citet{mrr89},  \citet{helou86}  created a parameterization of the
extragalactic population based on the \textit{IRAS} infrared colours in four bands. Within this framework, a
normal galaxy  component was defined by cool 100$\mu$m/60$\mu$m colours, a
starburst component defined by warm 100$\mu$m/60$\mu$m colours extending to an
ultraluminous component at higher infrared luminosities and an AGN component
was defined by high 25$\mu$m/60$\mu$m colours due to strong emission from the
mid-infrared  dust torus.
Two decades after the \textit{IRAS} all-sky survey, the Japanese \textit{AKARI} mission (launched successfully on Februrary 21st 2006, formerly known as \textit{ASTRO-F}) will  perform a new All-Sky Survey in 4 far-infrared and 2 mid-infrared bands to much improved sensitivity, higher spatial resolution and wider wavelength coverage  \citep{shib04,cpp04, cppshibai07}. AKARI is equipped with a 68.5 cm cooled telescope \citep{kaneda05} and two focal plane instruments, the Far-Infrared Surveyor (FIS) and the InfraRed Camera (IRC). The FIS has two 2-dimensional detector arrays and observes in four far-infrared bands between 50 and 180 $\mu$m (centred on N60 (65$\mu$m), WIDE-S (90$\mu$m),  WIDE-L (140$\mu$m), N160  (160$\mu$m), \citet{kawada04} ) . The IRC consists of three cameras covering 1.7-- 26 $\mu$m in 9 bands (N2  (2.4$\mu$m), N3  (3.2$\mu$m), N4  (4.1$\mu$m), S7  (7$\mu$m), S9W  (9$\mu$m),  S11 (11$\mu$m), L15 (15$\mu$m), L18W (18$\mu$m), L24  (24$\mu$m)) with fields of view of approximately 10$^{\prime}$  $\times$ 10$^{\prime}$ \citep{onaka04}. Both the FIS and IRC instruments also have capabilities for spectroscopy.
In addition to the All-Sky Survey, \textit{AKARI}  will also perform deep pointed surveys over areas of one to ten square degrees at the North (NEP) and South (SEP) Ecliptic Poles \citep{matsuhara06}. These major legacy surveys are expected to yield hundreds of thousands of sources and thus require a major data reduction effort.
As a precursor to these surveys, the aim of this work is to investigate the feasibility of colour segregation of the infrared populations with the minimum peripheral data and effort as possible in a similar manner to the \textit{IRAS} survey. Our colour segregation criteria will have merits not only for the \textit{AKARI}  data sets but also for similar large data sets expected from the  \textit{Spitzer} \&  \textit{Herschel} missions. In this paper we concentrate on the deep region selected for the \textit{AKARI} SEP survey while a more detailed investigation of our methods and their applications will be reported in  \citet{cppjeong07}.

\section{Phenomenological Model for Colour Segregations}

\subsection{Spectral Energy Distribution Library}

\begin{figure*}[ht]
  \begin{center}
    \epsfxsize = 15cm
    \epsffile{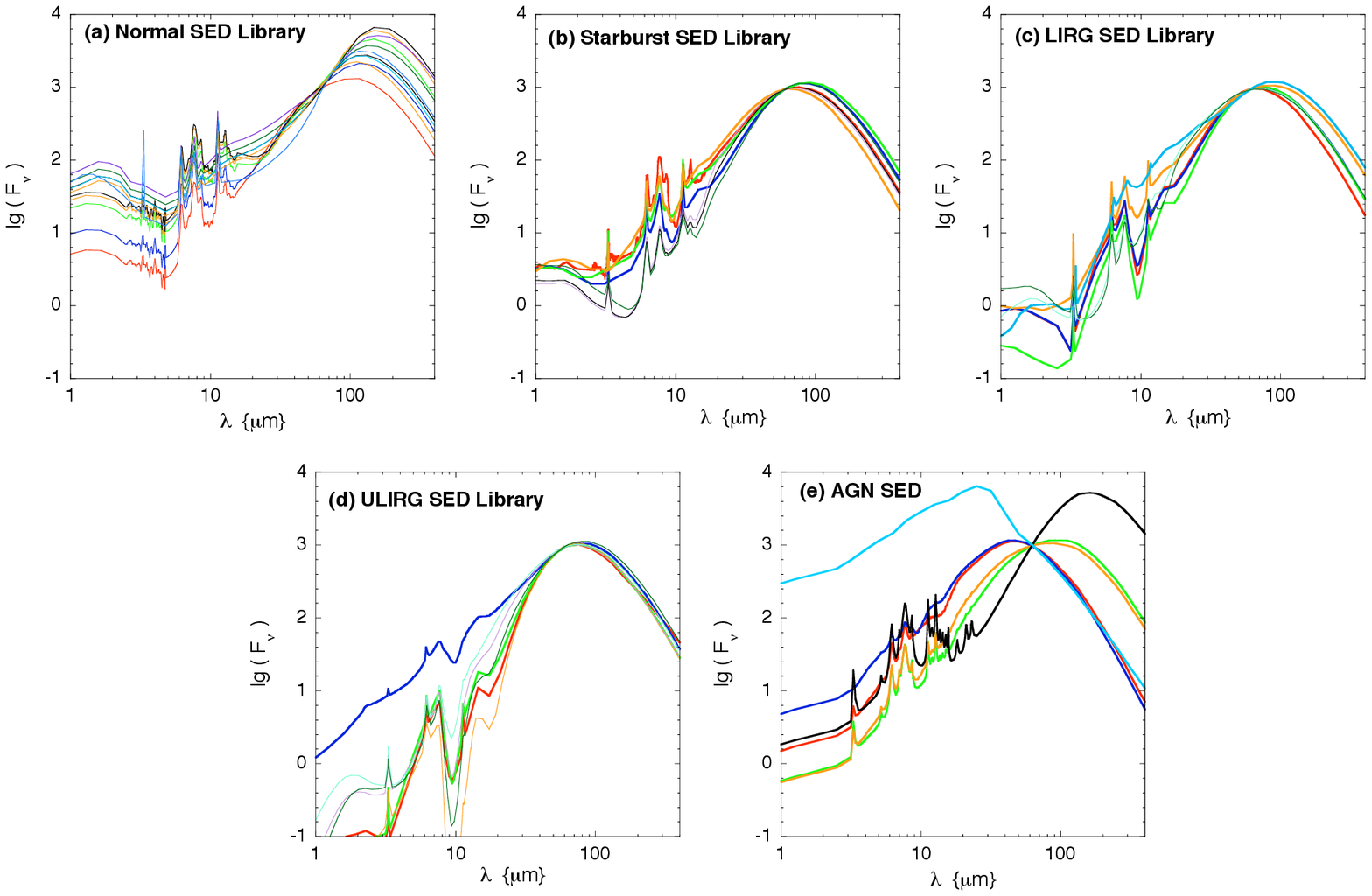}
    \end{center}
   \caption{ Selected spectral energy distributions (SED) from the spectral libraries described in the text  normalized to an arbitrary flux at 60$\mu$m. {\bf (a)}  Selected normal galaxies from the libraries of Efstathiou and Dale {\bf (b)} Selected starburst galaxies from the libraries of Efstathiou and Takagi. {\bf (c)} Selected LIRG sources from the libraries of Efstathiou and Takagi. {\bf (d)} Selected ULIRG sources from the libraries of Efstathiou and Takagi.  {\bf (e)} Selected AGN from the libraries of Siebenmorgen and Efstathiou.}
   \label{sed}
\end{figure*}

To model the colours of the extragalactic populations we have gathered a selection of spectral energy distribution (SED) templates for various galaxy populations from four large readily available contemporary spectral libraries. Computational constraints from our simulations restricts us from utilizing entire sets of thousands of SEDs so rather we have selected a moderate sub-sample which spans the entire range of spectral properties of the libraries. The selected sets of SEDs specifically model the quiescent normal galaxy population, the starburst population (increasing in luminosity: starburst, luminous infrared galaxies (LIRG), ultra-luminous infrared galaxies (ULIRG)) and the AGN population.
To model the quiescent normal galaxy population we have selected spectral templates from the libraries described in \citet{efstathiou03} and \citet{dale01}. The models of  \citet{efstathiou03} provide good fits to the \textit{IRAS}, \textit{ISO} and \textit{Spitzer} selected samples of the cool, quiescent galaxy population \citep{mrr04, mrr05} as well as the cold component of the sub-millimetre source population \citep{efstathiou03} and the galaxy source counts from sub-millimetre to near-infrared wavelengths \citep{cpp01,cpp07}. The model templates of  Dale et al.  have been shown to provide good fits to local normal galaxies from both  \textit{ISO} and more recently  \textit{Spitzer} selected samples \citep{dale01,dale05}, high-z Spitzer sources \citep{appleton04} and the \textit{ISO} \& \textit{Spitzer} mid-infrared source counts \citep{cpp05}. The starburst, LIRG, ULIRG population SEDs are taken from the radiative transfer models of  \citet{efstathiou00} and  \citet{takagi03a,takagi03b}. The starburst models of Efstathiou have provided good fits to a wide variety of individual objects from starburst (e.g. M82), LIRG (e.g.NGC6090,  IRAS1445-4343 ) to ULIRG sources (e.g. ARP220, MK231), \citep{efstathiou00,efstathiou05} and also the recent results from the  \textit{Spitzer} SWIRE survey \citep{mrr05}. The models of Takagi et al. provide good fits to the SEDs of dusty submillimetre galaxies \citep{takagi04}. The AGN population is modelled on the recent SED library of \citet{siebenmorgen04, siebenmorgen05}.  The SED libraries provide models for both Type 1 and Type 2 AGN   by the consideration of tapered disc geometries. 
The selected SEDs are shown in Figure \ref{sed}. Further details of the model selection are provided in   \citet{cppjeong07}

\subsection{Simulated Catalogues and Galaxy Colours}

To estimate the expected fluxes and corresponding colours in the \textit{AKARI} bands we use the contemporary galaxy evolution model of  \citet{cpp07}. This model uses a backward evolution process based on the  the \textit{IRAS} all-sky PSCz multi-component luminosity function defined at 60$\mu$m comprising of cool normal quiescent galaxies, and a warmer component defined by infrared luminosity as  $L_{IR}<10^{11}L_{\bigodot}$ starburst galaxies,  $L_{IR}>10^{11}L_{\bigodot}$ LIRG sources, $L_{IR}>10^{12}L_{\bigodot}$ ULIRG sources and AGN. This model fits the observed source counts from 2-1200$\mu$m from  the \textit{IRAS}, \textit{ISO}, \textit{Spitzer} missions and the SCUBA/JCMT, MAMBO/IRAM instruments  \citep{bertin97, pug99,dole04,fazio04,aussel99,serjeant00,papovich04,blain02,greve04}.  Our source colour information is simulated by passing a catalogue of point sources through the instrument simulator of \citet{jeong04}. An input catalogue is constructed for the 5 broad galaxy types outlined above within the framework of  \citet{cpp07}. The catalogue is constructed as a function of increasing redshift in incremental redshift bins. For each redshift bin, the theoretical number-redshift distribution from the model is used to randomly select the appropriate number of galaxies at that redshift and a corresponding luminosity  by sampling the input luminosity function. Each source from the catalogue is assigned a SED from the library depending on the galaxy type and the expected flux for each source from the corresponding SED at the desired wavelength and redshift is calculated and convolved with the appropriate filter response function.  A one sigma variation in the SED flux (i.e. for a given source in the catalogue the assigned flux at a given wavelength is selected randomly between $\pm$1$\sigma$ of the original SED template flux) is also included at this stage to allow for a wider range of output fluxes for a limited number of input SEDs.  Note that  although the instrument simulator is optimized for the \textit{AKARI} mission \citep{jeong04}, it is a simple matter to extend the simulator to encompass the characteristics of the infrared pass-bands of other space missions (e.g. \textit{Spitzer}, \textit{Herschel}, etc., see \citet{cppjeong07}). The final output from the simulator is a catalogue of sources with associated SED, and filter band convolved fluxes at a set of desired wavelengths.

\section{Predicted Colours of the Extragalactic Population in the AKARI SEP survey }

In this work, the simulated source catalogue is used to investigate the deep \textit{AKARI} SEP survey. The \textit{AKARI} SEP survey is a guaranteed time  \textit{AKARI} Mission Program (MP). The current projected area spans $\sim$8 square degrees  in a low cirrus brightness region centred at R.A.=4$^{h}$44$^{m}$00$^{s}$, Dec=-53deg 20$^{\prime}$00.$^{\prime \prime}$0 (J2000) where the cirrus brightness is spectacularly low $< B_{100} > \sim 0.2$ MJy/sr \citep{schlgel98}. This survey is estimated to reach the source confusion limit in the far-infrared  FIS wide bands and will be the deepest far-infrared image of the Universe to date (note that \citet{jeong06} predict confusion limits of $\sim$10mJy \& 60mJy in the FIS 90$\mu$m and 140$\mu$m bands respectively based on the source confusion criterion of 20 beams per source). Shallow mid-infrared data will also be available over the entire area, along with deep optical $R$-band imaging to 25th magnitude. In addition, an area of approximately 1 square degree will be covered by IRC mid-infrared imaging to $\sim$100$\mu$Jy. Source confusion at optical wavelengths may also pose complications for the identification of faint infrared sources. In the optical R-band, the source density may be up to an order of magnitude higher than the far-infrared AKARI source density. Furthermore, although the advent of wide field optical multi-object spectroscopy instruments, such as AAOmega on the Anglo-Australian Telescope \citep{sharp06} will make large scale follow up a reality, note that at these optical magnitudes (B$<$23) there will still be $\sim$5 optical sources per AKARI beam. Therefore, any identification / population segregation that is possible using the infrared data alone will be of great value.

\begin{figure*}[ht]
  \begin{center}
    \epsfxsize = 13cm
    \epsffile{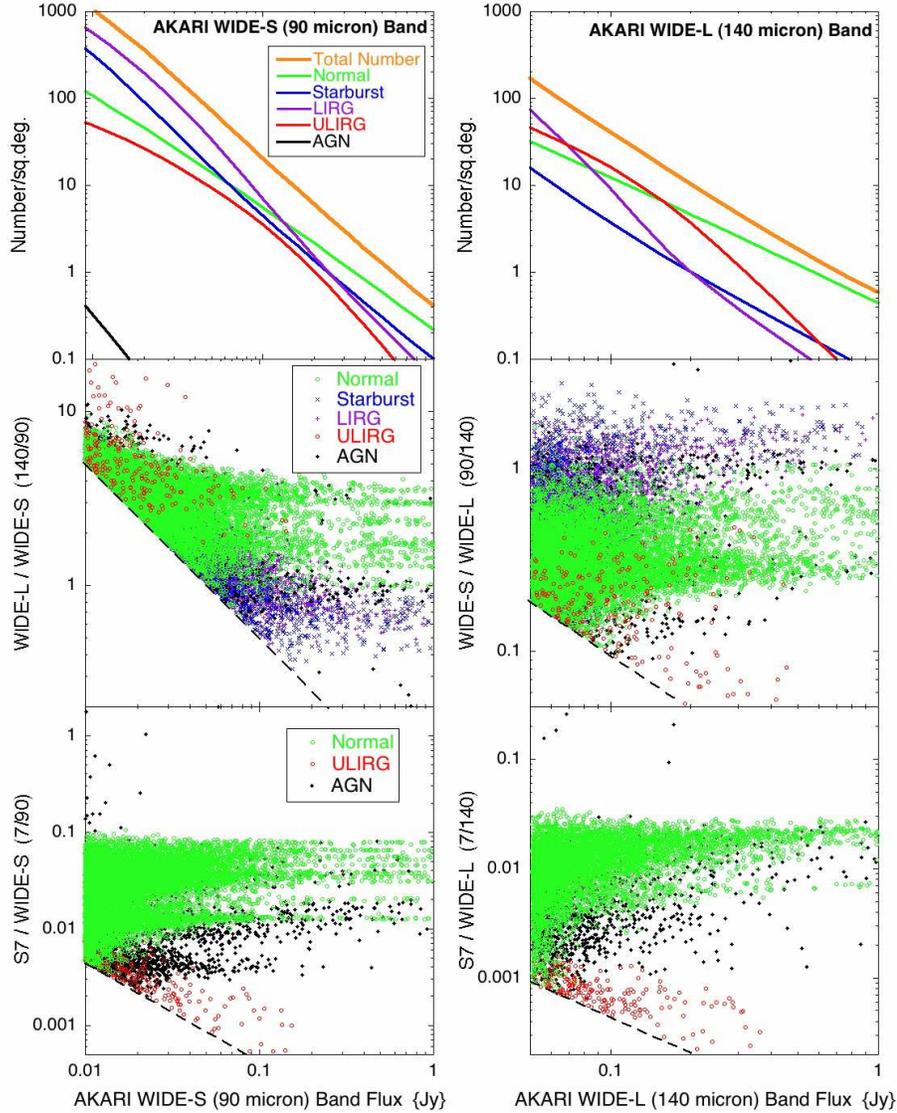}
    \end{center}
   \caption{ Example of colour segregation criteria. \textit{Top panel} shows the predicted source counts over 10mJy--1Jy \& 50mJy--1Jy from the models of \citet{cpp07} in the \textit{AKARI} FIS  WIDE-S 90$\mu$m (\textit{left})  \& WIDE-L 140$\mu$m (\textit{right}) bands respectively. Results are shown for total counts and component counts (Normal, starburst, LIRG, ULIRG, AGN populations). \textit{Middle panels} show an example colour criterion utilizing the \textit{AKARI} WIDE bands. The 140$\mu$m/90$\mu$m and 90$\mu$m/140$\mu$m colours are plotted as a function of 90$\mu$m \& 140$\mu$m flux respectively on the same flux scale as the source counts above.  \textit{Bottom panel} shows the segregation of the normal and ULIRG/AGN populations when one of the deep mid-infrared bands is used (IRC S7 7$\mu$m band) in conjunction with the FIS bands. In the figure, the \textit{green lines} and \textit{green open circles} represent the normal galaxies, the \textit{blue lines} and \textit{blue crosses} the starburst galaxies, \textit{pink lines} and \textit{pink plus symbols} the LIRG sources, \textit{red lines} and \textit{red open circles} the ULIRG sources and the \textit{black lines} and \textit{black filled circles} the AGN. The \textit{diagonal black dashed lines} in the bottom 4 panels correspond to the flux limits for the SEP survey in the relevant \textit{AKARI} band.}
   \label{counts}
\end{figure*}

In Figure \ref{counts}  the results of our analysis are presented. The \textit{top panel} shows the predicted source counts over 10mJy--1Jy \& 50mJy--1Jy from the models of \citet{cpp07} in the \textit{AKARI} FIS 90$\mu$m (\textit{left})  \& 140$\mu$m (\textit{right}) bands respectively. The \textit{AKARI}  SEP survey should probe to depths of $\sim$10mJy $\&$ 50mJy  (3$\sigma$) in the 90$\mu$m \& 140$\mu$m bands respectively. In the 90$\mu$m band, at brighter fluxes ($S>$100mJy) the source counts are predominantly dominated by cool normal galaxies due to the dust emission hump peaking in the range $\sim$100-200$\mu$m. At fainter fluxes  ($S<$100mJy) the warmer evolving starburst and LIRG populations dominate.  The 140$\mu$m band, samples the cooler Rayleigh-Jeans regime of the dust emission hump and is dominated by cool normal galaxies and cold ULIRG sources to $S\sim$100mJy. The fainter fluxes again see the emergence of the warmer LIRG population. 

The  \textit{middle panels} of Figure \ref{counts} show an example colour criterion utilizing the \textit{AKARI} WIDE bands. The 140$\mu$m/90$\mu$m and 90$\mu$m/140$\mu$m colours are plotted as a function of 90$\mu$m \& 140$\mu$m flux respectively on the same flux scales as the source counts in the \textit{top panel} in the figure. Limits in the flux and the colour corresponding to a 3$\sigma$ 90$\mu$m flux limit of 10mJy and a 140$\mu$m flux limit of 50mJy are plotted as  \textit{dotted} lines. Immediately it can be seen that the far-infrared colours provide a reasonable discriminator of galaxy type with the more \textit{active} star-forming galaxies occupying the parameter space corresponding to warmer colours (higher 90$\mu$m flux corresponding to 140$\mu$m/90$\mu$m colours $\lesssim$ 1.5 and 90$\mu$m/140$\mu$m colours $\gtrsim$ 1). Note that \citet{frayer06} have carried out a similar exercise with sources from the \textit{Spitzer} First Look Survey (FLS) although their approach was instead to fit different dust temperatures to the far-infrared colours using the MIPS 160$\mu$m and 70$\mu$m bands, interpreting a cooler dust temperature  as cold infrared colours. \textit{AKARI}, with its 4 FIS bands will have a greater power than \textit{Spitzer} in such far-infrared population colour segregation. In addition to providing a crude separation of cool normal galaxies and warmer starburst and LIRG sources, the FIS colour segregation is also capable of identifying a distinct cold ULIRG population.  These sources occupy a unique colour-flux parameter space in Figure \ref{counts}. For the case of the 140$\mu$m band (\textit{right-hand middle panel}), the ULIRGs occupy the parameter space defined by S(140$\mu$m)$\gtrsim$100mJy with 90$\mu$m/140$\mu$m colour $\lesssim$0.2. In the case of the 90$\mu$m band the ULIRGs are found at fainter fluxes of  S(90$\mu$m)$\lesssim$ 30mJy  with 140$\mu$m/90$\mu$m colours $\gtrsim$ 6. The models of \citet{cpp07} predict  that in the 90$\mu$m band, these colour selected ULIRGS will be high redshift (1$<$z$<$3) sources whilst the 140$\mu$m colour selected ULIRGs will be predominantly at redshifts z$<$1.5. The fainter 90$\mu$m ULIRG sample should include almost all the 140$\mu$m ULIRGs. 
From the \textit{middle panel} in Figure \ref{counts} it can also be seen that the ULIRG samples selected in both the 90$\mu$m \& 140$\mu$m bands  descend and ascend respectively  into the colour parameter space occupied by the normal galaxies, i.e. that there is contamination of the cold ULIRG population by the cooler colours in the normal galaxy population. This contamination of ULIRGs by cool (low redshift) normal galaxies has also been reported in other colour selection  studies over far-mid-infrared wavelengths \citep{takagi05}. Using suitable additional colour criteria it is also possible to de-convolve the ULIRG sample from the cool normal galaxy population. The  \textit{bottom panel} of Figure \ref{counts}  gives an example of the potential power of additional colour constraints to further segregate the galaxy populations. The  \textit{bottom panel} shows the segregation of the normal and ULIRG populations when one of the deep mid-infrared bands is used (in this case the IRC S7 7$\mu$m band) in conjunction with the FIS bands. The \textit{bottom left panel} shows the results for the 7$\mu$m/90$\mu$m colours as a function of 90$\mu$m flux. The normal galaxies are separated from the cold ULIRG population by virtue of their higher mid-infrared/far-infrared flux ratio (see Figure \ref{sed}). The slight remaining contamination that can be seen is actually from a hotter ULIRG population whose mid-infrared emission is dominated by an AGN. An even more significant segregation can be achieved using the 7$\mu$m/140$\mu$m colours as a function of the longer wavelength 140$\mu$m flux. Since  the far-infrared emission  of the cold ULIRGs is redshifted into the 140$\mu$m band, the mid-infrared/far-infrared flux ratio is even lower. In addition, the normal galaxies exhibit a smaller scatter in 7$\mu$m/140$\mu$m colours as a function of 140$\mu$m flux since there is less variation in the SEDs on the Rayleigh-Jeans slope than at the shorter wavelengths sampled by the 90$\mu$m band. The \textit{bottom panels} of Figure \ref{counts} also show that, although to a lesser extent, a similar segregation can also be achieved for the AGN population.

\section{Conclusions}

Using a suite of SED libraries combined with a contemporary galaxy evolution model and instrument simulator, we have carried out a pilot study of, and demonstrated a method to segregate galaxy populations in large data sets in a similar manner to the methods utilized for the \textit{IRAS} all-sky survey. Using simple colour criteria, the normal galaxy, starburst/LIRG \& ULIRG populations can be succesfully discriminated as a function of far-infrared flux without the need for large-scale optical follow up campaigns. This method has been applied to the \textit{AKARI} deep survey at the SEP where it is successful in achieving a general segregation of the various extragalactic populations. The addition of more bands with a greater dispersion in wavelength can be used to further refine and decontaminate specific target populations (in this example the ULIRG and normal galaxy populations).
The colour segregation criteria can thus be used to construct basic multi-component source counts without the need for expensive ground-based follow up. This is illustrated in  Figure \ref{counts}. For the  \textit{AKARI} FIS 90$\mu$m band, the normal galaxies are identified with sources having colours over the range 1.5$\lesssim$140$\mu$m/90$\mu$m$\lesssim$5 over all values of 90$\mu$m flux. The starburst/LIRG sources are identified with sources of colours 90$\mu$m/140$\mu$m colours $\gtrsim$ 1. The ULIRG sources can be identified by the dual colour criterea of 140$\mu$m/90$\mu$m$\gtrsim$6  \& 7$\mu$m/90$\mu$m$\lesssim$0.007. In the 140$\mu$m band, the normal galaxies can be identified with sources of colour 0.2$\lesssim$90$\mu$m/140$\mu$m$\lesssim$0.7 while the starburst \& LIRG sources have generally warmer 90$\mu$m/140$\mu$m colours $\gtrsim$0.7. The ULIRGs in the 140$\mu$m band are defined as sources of colour 90$\mu$m/140$\mu$m$\lesssim$0.2 \& 7$\mu$m/140$\mu$m$\lesssim$0.02  over a flux range S(140$\mu$m)$\gtrsim$ 100mJy. Although not the focus of this present work, a similar criterion can also be applied to the AGN population.

The colour-segregation method and corresponding colour segregated source counts holds great promise for the data sets that are expected from the large area surveys that are being conducted or will be conducted by the present and next generation of space-borne far-infrared telescopes such as  \textit{Spitzer},  \textit{AKARI} \&  \textit{Herschel}.
A more detailed description of this work and application to the  \textit{AKARI} All-Sky Survey and possible surveys with  \textit{Herschel} will be presented in \citet{cppjeong07}

\section{Acknowledgments}
The authors would like to thank the referees whose comments greatly improved the presentation of this paper. W.-S. Jeong acknowledges a JSPS fellowship to Japan.
The {\it AKARI} Project is an infrared mission of the Japan Space Exploration Agency (JAXA) Institute of Space and Astronautical Science (ISAS), and is carried out with the participation of mainly the following institutes; Nagoya University, The University of Tokyo, National Astronomical Observatory Japan, The European Space Agency (ESA), Imperial College London, University of Sussex, The Open University (UK), University of Groningen / SRON (The Netherlands), Seoul National University (Korea). The far-infrared detectors were developed under collaboration with The National Institute of Information and Communications Technology.


\newpage

\end{document}